# New limit for $\nu_e \to \nu_\tau$ oscillations

E. Arik[a], A. Mailov[a,b], S. Sultansoy[b,c]

[a]Dept. of Physics, Bogazici University, 80815 Bebek, Istanbul, TURKEY
[b]Institute of Physics, Academy of Sciences, H. Cavid Ave. 33, Baku, AZERBAIJAN
[c]Dept. of Physics, Ankara University, 06100 Tandogan, Ankara, TURKEY

**Abstract**

We present a calculation for a new limit on $\nu_e \to \nu_\tau$ oscillation based on the recent data from CHORUS experiment.

We are far from understanding the mass spectrum and mixings of fundamental fermions. For the leptonic sector, experimental data set only upper limits for neutrino masses, whereas the masses of charged leptons (especially $m_e$ and $m_\mu$) are measured with high precision. Unlike the quark sector, we do not have any established information on lepton mixings. If neutrinos have masses much less than the experimental upper bounds [1], then the most promising way to obtain the appropriate information is to search for neutrino oscillations.

Current accelerator based searches for neutrino oscillations are concentrated mainly on $\nu_\mu \to \nu_\tau$ options [2, 3]. However, if the recent results from Super-Kamiokande [4] really reflect the existence of $\nu_\mu \to \nu_\tau$ oscillations, then this can only be observed in experiments using long-baseline neutrino beams. Super-Kamiokande data yields $sin^2 2\theta_{\mu\tau} \approx 0.82$ and $\Delta m^2 = O(10^{-3}\ eV^2)$ but CHORUS and NOMAD experiments are sensitive to $\Delta m^2 = O(1\ eV^2)$ with $L \approx 1\ km$. A similar detector set up at $L \approx 10^3\ km$ will have the desired sensitivity. The idea of CERN long-baseline neutrino beam to Gran Sasso laboratory [5] seems to be very promising in observing $\nu_\mu \to \nu_\tau$ oscillations.

On the other hand, one can find a number of arguments in favour of $\nu_e \to \nu_\tau$ oscillations. For example, this is preferable in the framework of direct generalization [6] of the first lepton number scheme proposed by Konopinski-Mahmoud [7] and Zel'dovich [8]. The dominance of $\nu_e \to \nu_\tau$ oscillations is also predicted by several preonic models [9].

Due to the above reasons, a search for $\nu_e \to \nu_\tau$ oscillations becomes more important for the existing accelerator based experiments. Current limits on $\nu_e \to \nu_\tau$ oscillations are rather poor: $sin^2 2\theta_{e\tau} < 0.25$ for large $\Delta m^2$ [1, 10]. In this note, we show that the recent data from CHORUS improves this bound essentially, and the second phase of analysis (with larger statistics etc.) will lead to further improvement at least by an order.

For clearness let us consider the large $\Delta m^2$ values. In this case the expected number of observed $\tau^-$ decays for channel $i$ with branching ratio $Br_I$ is given by

$$N_{\tau_i}^{\nu_\mu} = \frac{1}{2} \cdot \sin^2 2\theta_{\mu\tau} \cdot Br_i \cdot \int \Phi_{\nu_\mu} \cdot \sigma_\tau \cdot A_{\tau_i}^{\nu_\mu} \cdot \varepsilon_{\tau_i}^{\nu_\mu} \cdot dE \qquad (1)$$

for $\nu_\mu \to \nu_\tau$ oscillations, and

$$N_{\tau_i}^{\nu_e} = \frac{1}{2} \cdot \sin^2 2\theta_{e\tau} \cdot Br_i \cdot \int \Phi_{\nu_e} \cdot \sigma_\tau \cdot A_{\tau_i}^{\nu_e} \cdot \varepsilon_{\tau_i}^{\nu_e} \cdot dE \qquad (2)$$

for $\nu_e \to \nu_\tau$ oscillations.

In Eqs.(1) and (2), $\Phi_{\nu_\mu}$ ($\Phi_{\nu_e}$) is the incident $\nu_\mu$ ($\nu_e$) flux spectrum, $\sigma_\tau$ is the cross-section of the charged $\nu_\tau$ interaction, $A_{\tau_i}^{\nu_\mu}$ ($A_{\tau_i}^{\nu_e}$) is the acceptance and reconstruction efficiency for the corresponding $\tau$ decay events originating from $\nu_\mu$ $\rightarrow \nu_\tau$ ($\nu_e \rightarrow \nu_\tau$) oscillations and $\varepsilon_{\tau_i}^{\nu_\mu}$ ($\varepsilon_{\tau_i}^{\nu_e}$) is the energy dependent kink detection efficiency.

For the CERN neutrino beam, average energy of $\nu_e$ is higher than the average energy of $\nu_\mu$, hence we can state that $(A_{\tau_i}^{\nu_\mu} \cdot \varepsilon_{\tau_i}^{\nu_\mu}) < (A_{\tau_i}^{\nu_e} \cdot \varepsilon_{\tau_i}^{\nu_e})$. Therefore, one can easily obtain the following relation

$$\sin^2 2\theta_{e\tau} = k \cdot \sin^2 2\theta_{\mu\tau} \cdot \frac{\int \Phi_{\nu_\mu} \sigma_\tau dE}{\int \Phi_{\nu_e} \sigma_\tau dE}$$
$$= k \cdot \sin^2 2\theta_{\mu\tau} \cdot \frac{<\sigma_\tau>_{\nu_\mu}}{<\sigma_\tau>_{\nu_e}} \cdot \frac{\int \Phi_{\nu_\mu} dE}{\int \Phi_{\nu_e} dE} \quad (3)$$

where $k < 1$. We can assume $<\sigma_\tau>_{\nu_\mu} / <\sigma_\tau>_{\nu_e} = <E_{\nu_\mu}> / <E_{\nu_e}>$ with good accuracy. Using results of the neutrino beam simulations for CHORUS [11]: $<E_{\nu_\mu}> \approx 26.6$ GeV, $<E_{\nu_e}> \approx 39.6$ GeV and $(\int \Phi_{\nu_e} dE)/(\int \Phi_{\nu_\mu} dE) = 8.5 \cdot 10^{-3}$, we obtain

$$\sin^2 2\theta_{e\tau} \approx 80 \cdot k \cdot \sin^2 2\theta_{\mu\tau}. \quad (4)$$

According to Eqn.(4), recent CHORUS upper limit on $\nu_\mu \rightarrow \nu_\tau$ oscillations, $sin^2 2\theta_{\mu\tau} < 1.2 \cdot 10^{-3}$ [12], simultaneously gives $sin^2 2\theta_{e\tau} < 0.10$ for $\nu_e \rightarrow \nu_\tau$ oscillations at large $\Delta m^2$, which is more than two times stringent compared to the previous limit [1]. Furthermore, designed sensitivity of CHORUS, namely $sin^2 2\theta_{\mu\tau} < 2 \cdot 10^{-4}$, will correspond to $sin^2 2\theta_{e\tau} < 0.016$. Moreover, combining results from CHORUS and NOMAD (because both experiments use the same neutrino beam), one can achieve additional improvement.

Finally, if $\tau^-$ events are observed by CHORUS and NOMAD, one will be able to distinguish between $\nu_e \rightarrow \nu_\tau$ and $\nu_\mu \rightarrow \nu_\tau$ oscillations using the transverse position distribution of $\tau^-$ events, because $\nu_e$ beam distribution is approximately flat whereas $\nu_\mu$ beam has a bell shaped distribution with a maximum at $r = 0$ [3].

In conclusion, we have shown that the CHORUS and NOMAD results on $\nu_\mu \to \nu_\tau$ oscillations give possibility to obtain the upper limit for $\nu_e \to \nu_\tau$ oscillations too. Our estimations are approximate ones, and the calculated upper limit for $sin^2 2\theta_{e\tau}$ is rather conservative. More detailed calculations will further improve this limit by $\approx$ 10÷20 %.

# References

[1] Review of Particle Physics, The European Physical Journal **C3** (1998) N1-4.

[2] CHORUS Collab., E. Eskut et al., Nucl. Inst. and Meth. **A401** (1997) 7.

[3] NOMAD Collab., J. Altegoer et al., Nucl. Inst. and Meth. **A404** (1998) 96.

[4] Super-Kamiokande Collab., Y. Fukuda et al., hep-ex/9807003.

[5] G. Acquistapace et al., CERN 98-02, INFN/AE-98/05 (1998).

[6] B.A. Arbuzov and S.F. Sultanov, Sov. J. Nucl. Phys. **33** (1981) 223.

[7] E.J. Konopinski and H.M. Mahmoud, Phys. Rev. **92** (1953) 1045.

[8] Y.B. Zel'dovich, Doklady Akad. Nauk SSSR **91** (1953) 1317.

[9] J.J. Dugne et al., hep-ph/9802339.

[10] E531 Collab., N. Ushida et al., Phys. Rev. Lett. **57** (1986) 2897.

[11] S. Sorrentino, CHORUS Internal Note 98005 (1998).

[12] CHORUS Collab., contributed paper at the XXIX Int. Conf. On High Energy Physics, 23-28 july, 1998, Vancouver, BC, Canada.